\documentstyle{cjp3}

\begin{document}

\title{A Brief Introduction to Chiral Perturbation Theory}

\authori{Barry R. Holstein}

\addressi{Department of Physics and Astronomy, University of Massachusetts, 
Amherst, MA 01003, USA}

\authorii{}       
\addressii{}
\authoriii{}      
\addressiii{}     

\headtitle{Chiral Perturbation Theory}
\headauthor{Barry R. Holstein}

\evidence{A}
\daterec{XXX}    
\cislo{0}  \year{1999}
\setcounter{page}{1}

\maketitle
\begin{abstract}
A brief introduction to the subject of chiral perturbation theory ($\chi$pt) 
is presented, including
a discussion of effective field theory and applications of $\chi$pt 
in the arena of purely mesonic interactions as well as in the $\pi$N 
sector.     
\end{abstract}
\section{Introduction}
For the past three decades or so the holy grail 
sought by particle/nuclear knights has been to verify the correctness of 
the ``ultimate''
theory of strong interactions---quantum chromodynamics (QCD).  
The theory is, of course,
deceptively simple on the surface.
Indeed the form of the Lagrangian\footnote{Here the covariant derivative is
\begin{equation}
i D_{\mu}=i\partial_{\mu}-gA_\mu^a {\lambda^a \over 2} \, ,
\end{equation}
where $\lambda^a$ (with $a=1,\ldots,8$) are the SU(3) Gell-Mann matrices,
operating in color space, and the color-field tensor is defined by
\begin{equation}
G_{\mu\nu}=\partial_\mu  A_\nu -  \partial_\nu  A_\mu -
g [A_\mu,A_\nu]  \, ,
\end{equation} }
\begin{equation}
{\cal L}_{\mbox{\tiny QCD}}=\bar{q}(i  {\not\!\! D} - m )q-
{1\over 2} {\rm tr} \; G_{\mu\nu}G^{\mu\nu} \, .
\end{equation}
is elegant, and the theory is renormalizable.  So why are we still not
satisfied?  While at the very largest energies, asymptotic freedom 
allows the use of perturbative
techniques, for those who are interested in making contact with low energy 
experimental findings there exist at least three fundamental difficulties:
\begin{itemize}
\item [i)] QCD is written in terms of the "wrong" degrees of 
freedom---quarks and
gluons---while low energy experiments are performed with hadronic bound states;

\item [ii)] the theory is non-linear due to gluon self interactions;

\item[iii)] the theory is one of strong coupling---$g^2/4\pi\sim 1$---so that 
perturbative methods are not practical.
\end{itemize}
Nevertheless, there has been a great deal of recent progress in 
making contact between
theory and experiment using the technique of "effective field theory'', 
which exploits the 
chiral symmetry of the QCD interaction.  In order to understand how 
this is accomplished, we
first review symmetry breaking as well as the concept of effective 
interactions.  Then
we show how these ideas can be married via chiral perturbation theory 
and indicate a few
contemporary physics applications.

\section{Symmetry and Symmetry Breaking}

The importance of symmetry in physics is associated with Noether's 
theorem, which states that associated with any symmetry in physics is a 
corresponding
conservation law.  Thus, for example, 
\begin{itemize}
\item [i)] translation invariance implies conservation of momentum;
\item [ii)] time translation invariance implies conservation of energy;
\item [iii)] rotational invariance implies conservation of angular momentum
\end{itemize} 
These, however, are perhaps the only exact symmetries in nature.  All others
are broken in some way and there are in general only three types of symmetry
breaking which can occur:

{\it Explicit}:  The most familiar is {\it explicit} symmetry 
breaking, wherein the breaking 
occurs in the Lagrangian itself.  As an example, consider the 
harmonic oscillator Lagrangian
\begin{equation}
L={1\over 2}m\dot{x}^2-{1\over 2}m\omega^2x^2
\end{equation}
which is clearly symmetric under spatial inversion---$x\rightarrow -x$.  
Correspondingly, the ground state
(lowest energy) configuration---$x=0$, which is found via ${\partial L\over 
\partial x}=0$---also shares this symmetry.  On the other hand, if we add a
linear potential (constant force) into the system, the Lagrangian becomes
\begin{equation}
L={1\over 2}m\dot{x}^2-{1\over 2}m\omega^2x^2+\lambda x
\end{equation}
The ground state is now given by $x=\lambda/m\omega^2$, which no longer
is invariant under spatial inversion, but this is to be expected because of the
{\it explicit} symmetry breaking term---$\lambda x$---in the Lagrangian.

{\it Spontaneous}:  Less familiar but still relatively common 
is {\it spontaneous} symmetry
breaking, wherein the Lagrangian of a system possesses a symmetry 
but the ground
state does not.  A simple classical physics example of this is the case of
a thin hoop of radius $R$ immersed in a gravitational field and 
rotating about a vertical axis at fixed angular velocity $\omega$.  
A bead which can move without friction along the hoop is then described by 
the Lagrangian\cite{ssb}
\begin{equation}
L={1\over 2}m(R^2\dot{\theta}^2+\omega^2R^2\sin^2\theta)+mgR\cos\theta
\end{equation}
where $\theta$ is the angle subtended by the bead from the bottom of the hoop.
Clearly the Lagrangian is invariant under 
reflection---$L(\theta)\rightarrow L(-\theta)$---but the ground state 
configuration is given by
\begin{equation}
{\partial L\over \partial \theta}=m\omega^2R^2\sin\theta(\cos\theta-{g\over 
m\omega^2R})=0,
\end{equation}
which has the stable equilibirium solution $\theta=+\cos^{-1}{g\over 
\omega^2R}$ {\it or} $\theta=-\cos^{-1}{g\over \omega^2R}$ 
if ${g\over \omega^2R}<1$.
Obviously in this situation the ground state breaks the symmetry under
$\theta\rightarrow -\theta$, even
though the Lagrangian does not---this is an example of {\it spontaneous} 
symmetry breaking.

{\it Anomalous}:  Finally, we consider {\it anomalous} or quantum 
mechanical symmetry breaking
wherein the Lagrangian at the classical level is symmetric, but the 
symmetry is broken upon quantization.  Obviously there are no classical
physics examples of this phenomenon and, to my knowledge, every manifestation
except one is in the arena of quantum field theory.  The one example from
ordinary quantum mechanics involves the breaking of scale invariance by
a two-dimensional delta function potential\cite{afp}.  
To set the stage, first 
consider a {\it free} particle of mass $m$, which satisfies the time-independent 
Schrodinger equation
\begin{equation}
-{1\over 2m}\nabla^2\psi={k^2\over 2m}\psi
\end{equation}
A general partial wave solution can be written as
\begin{equation}
\psi_\ell(\vec{r})={1\over r}\sum_\ell a_\ell\chi_\ell(r)P_\ell(\cos\theta)
\end{equation}
where $\chi(r)$ satisfies the differential equation
\begin{equation}
\left(-{d^2\over dr^2}+{\ell(\ell+1)\over r^2}+k^2\right)\chi_\ell(r)=0
\end{equation}
Obviously there exists a scale invariance here---the Schrodinger equation is
invariant under the scale transformation $r\rightarrow\lambda r,k
\rightarrow k/\lambda$, a consequence of which is that the solution must be
a function only of $k$ {\it times } $r$ and not of $k$ or $r$ alone.  
For example, a free particle solution can be
written in the form
\begin{equation}
\psi(\vec{r})=e^{ikz}=e^{ikr\cos\theta}\stackrel{r\rightarrow\infty}{\longrightarrow}
{1\over 2ikr}\sum_{\ell}
(2\ell+1)P_\ell(\cos\theta)(e^{ikr}-e^{-i(kr-\ell\pi)})
\end{equation} 
Note that there exists a phase shift $\ell\pi$ of the outgoing spherical 
wave with respect to its incoming counterpart.  This phase shift is, however,
$k-independent$ as required by scale invariance.  On the other hand if we
include a potential $V(r)$ then the solution 
has the asymptotic form
\begin{equation}
\psi^{(+)}(\vec{r})\stackrel{r\rightarrow\infty}{\longrightarrow}
e^{ikz}+{e^{ikr}\over r}f_k(\theta)
\end{equation} 
where the scattering amplitude $f_k(\theta)$ is given by 
\begin{equation}
f_k(\theta)=\sum_{\ell}(2\ell+1){e^{2i\delta_\ell(k)}-1\over 2ik}
P_\ell(\cos\theta)
\end{equation}
In this case there exists an additional phase shift $\delta_\ell(k)$ in each 
partial wave, which breaks the scale invariance, but this is to be expected
because of the presence of the (symmetry violating) potential.

In the case of two dimensions, one can write the scattering wave function
in the asymptotic form
\begin{equation}
\psi^{(+)}(\vec{r})\stackrel{r\rightarrow\infty}{\longrightarrow}
e^{ikz}+{1\over \sqrt{r}}
e^{i(kr+{\pi\over 4})}f_k(\theta)
\end{equation}
with scattering amplitude
\begin{equation}
f_k(\theta)=-i\sum_{m=-\infty}^\infty{e^{2i\delta_m(k)}-1\over \sqrt{2\pi k}}
e^{im\theta}
\end{equation}
where now we expand in terms of exponentials $e^{im\theta}$ instead of Legendre
polynomials.  What is special about two dimensions is that it is possible to
introduce a {\it scale invariant} potential $V(\vec{r})=g\delta^2(\vec{r})$.  
The associated differential cross section is found to be
\begin{equation}
{d\sigma\over d\Omega}\stackrel{k\rightarrow\infty}{\longrightarrow}
{\pi\over 2k\ln({k^2\over \mu^2})}
\end{equation} 
which corresponds to pure $m=0$ scattering with an energy dependent phase
shift 
\begin{equation}
\cot\delta_0(k)={1\over \pi}\ln{k^2\over \mu^2}-{2\over g}
\end{equation}
The scale invariance present at the classical level (no scattering cross 
section since we have a delta function potential) is then 
violated upon quantization.

Interestingly, QCD makes use of all three forms of symmetry breaking!

\section{Effective Field Theory}

The power of effective field theory is associated with the feature that there
exist many situations in physics involving two scales, one heavy and one 
light.  Then if one is working at energies small compared to the heavy scale, 
one can fully describe the interactions in terms of an ``effective'' picture,
which is written only in terms of the light degrees of freedom, but which
fully includes the influence of the heavy mass scale through  
virtual effects.  A number of good review articles exist concerning
the subject\cite{eftr}.
 
Before proceeding to QCD, however, it is useful to study this idea 
in the simpler context of ordinary quantum mechanics, in order to get familiar 
with the concept.
Specifically, we examine the question of why the sky is blue, whose 
answer can be
found in an analysis of the scattering of photons from the sun 
by atoms in the atmosphere---Compton scattering\cite{skb}.  First we examine 
the problem using traditional quantum mechanics and consider 
elastic (Rayleigh) scattering from (for simplicity)
single-electron (hydrogen) atoms.  The appropriate Hamiltonian is then
\begin{equation}
H={(\vec{p}-e\vec{A})^2\over 2m}+e\phi
\end{equation}
and the leading---${\cal O}(e^2)$---amplitude for Compton scattering
is given by the Kramers-Heisenberg form
\begin{eqnarray}
{\rm Amp}&=&-{e^2/m\over \sqrt{2\omega_i2\omega_f}}\left[\hat{\epsilon}_i\cdot
\hat{\epsilon}_f^*+{1\over m}\sum_n\left({\hat{\epsilon}_f^*\cdot
<0|\vec{p}e^{-i\vec{q}_f\cdot\vec{r}}|n>
\hat{\epsilon}_i\cdot <n|\vec{p}e^{i\vec{q}_i\cdot\vec{r}}|0>\over 
\omega_i+E_0-E_n}\right.\right.
\nonumber\\
&+&\left.\left.{\hat{\epsilon}_i\cdot <0|\vec{p}e^{i\vec{q}_i\cdot\vec{r}}|n>
\hat{\epsilon}_f^*\cdot <n|\vec{p}e^{-i\vec{q}_f\cdot\vec{r}}|0>\over E_0-\omega_f-E_n}
\right)\right]
\end{eqnarray}
where $|0>$ represents the hydrogen ground state having binding energy
$E_0$. 

Here the leading component is the familiar 
$\omega$-independent Thomson amplitude and would appear naively 
to lead to an energy-independent
cross-section.  However, this is not the case.  Indeed, by expanding 
in powers of $\omega$ and using a few quantum mechanical tricks, then
provided that the energy of the photon is much smaller than a typical 
excitation energy---as is the case for optical photons---the cross section 
can be written as
\begin{eqnarray}
\quad {d\sigma\over d\Omega}&=&
\lambda^2\omega^4|\hat{\epsilon}_{f}^*\cdot
\hat{\epsilon}_{i}|^2\left(1+{\cal O}\left({\omega^2\over (\Delta E)^2}
\right)\right)\label{eq:cc}
\end{eqnarray}
where 
\begin{equation}
\lambda=\alpha_{em}\sum{2|z_{n0}|^2\over E_n-E_0}
\end{equation}
is the atomic electric polarizability,
$\alpha_{em}=e^2/4\pi$ is the fine structure constant, and $\Delta E\sim m
\alpha_{em}^2$ is a typical hydrogen excitation energy.  We note that  
$\alpha_{em}\lambda\sim a_0^2\times {\alpha_{em}\over \Delta E}\sim a_0^3$ 
is of order the atomic volume, as will be exploited below, and that the cross 
section itself has the characteristic $\omega^4$ 
dependence which leads to the blueness of the sky---blue light scatters much
more strongly than red\cite{feyn}.

Now while the above derivation is certainly correct, it requires somewhat detailed and
lengthy quantum mechanical manipulations which obscure the relatively simple physics
involved.  One can avoid these problems by the use of effective field theory 
methods.  The key point is that of scale.  
Since the incident photons
have wavelengths $\lambda\sim 5000$A much larger than the $\sim$ 1A atomic size, then at 
leading order the photon is insensitive to the presence of the atom, since the latter
is electrically neutral.  If $\chi$ represents the wavefunction of the atom,
then the 
effective leading order Hamiltonian is simply  
\begin{equation}
H_{eff}^{(0)}=\chi^*\left({\vec{p}^2\over 2m}+e\phi\right)\chi
\end{equation}
and there is {\it no} interaction with the field.  In higher orders, 
there {\it can} exist such atom-field interactions and this is
where the effective Hamiltonian comes in to play.  In order to construct the
effective interaction, we demand certain general principles---this
Hamiltonian must satisfy fundamental symmetry requirements.  In
particular $H_{eff}$ must be gauge invariant, must be a scalar under
rotations, and must be even under both parity and time reversal 
transformations.  Also,
since we are dealing with Compton scattering, $H_{eff}$ should be
quadratic in the vector potential.
Actually, from the requirement of gauge invariance it is
clear that the effective interaction can utilize only the electric and magnetic
fields
\begin{equation}
\vec{E}=-\vec{\nabla}\phi-{\partial\over \partial t}\vec{A}, 
\qquad \vec{B}=\vec{\nabla}\times\vec{A}\label{eq:ii}
\end{equation}
since these are invariant under a gauge transformation
\begin{equation}
\phi\rightarrow\phi+{\partial\over \partial t}\Lambda,\qquad \vec{A}
\rightarrow\vec{A}-\vec{\nabla}\Lambda
\end{equation}
while the vector and/or scalar potentials are not.  The lowest order
interaction then can involve only the rotational invariants 
$\vec{E}^2,\vec{B}^2$
and $\vec{E}\cdot\vec{B}$.  However, under spatial
inversion---$\vec{r}\rightarrow -\vec{r}$---electric and magnetic
fields behave oppositely---$\vec{E}\rightarrow -\vec{E}$ while
$\vec{B}\rightarrow\vec{B}$---so that parity invariance rules out any
dependence on $\vec{E}\cdot\vec{B}$.  Likewise under time
reversal---$t\rightarrow -t$---we have $\vec{E}\rightarrow \vec{E}$ but
$\vec{B}\rightarrow -\vec{B}$ so such a term is also ruled out by time
reversal invariance.  The simplest such effective Hamiltonian must
then have the form
\begin{equation}
H_{eff}^{(1)}=\chi^*\chi[-{1\over 2}c_E\vec{E}^2
-{1\over 2}c_B\vec{B}^2]\label{eq:ll}
\end{equation}
(Terms involving time or spatial derivatives are much smaller.)
We know from electrodynamics that 
${1\over 2}(\vec{E}^2+\vec{B}^2)$
represents the field energy per unit volume, so by dimensional
arguments, in order to represent an
energy in Eq. \ref{eq:ll}, $c_E,c_B$ must have dimensions of volume.
Also, since the photon has such a
long wavelength, there is no penetration of the atom, so  only classical scattering
is allowed.  The relevant scale must then be atomic size so that we can write
\begin{equation}
c_E=k_Ea_0^3,\qquad c_B=k_Ba_0^3
\end{equation}
where we anticipate $k_E,k_B\sim {\cal O}(1)$.  Finally, since for photons
with polarization $\hat{\epsilon}$ and four-momentum $q_\mu$ we
identify $\vec{A}(x)=\hat{\epsilon}\exp(-iq\cdot x)$,
then from Eq. \ref{eq:ii}, $|\vec{E}|\sim \omega$, 
$|\vec{B}|\sim |\vec{k}|=\omega$ and 
\begin{equation}
{d\sigma\over d\Omega}\propto|<f|H_{eff}|i>|^2\sim\omega^4 a_0^6
\end{equation}
as found in the previous section via detailed calculation.  Clearly
the effective interaction method provides and efficient and insightful
way in which to perform the calculation.

\section{Application to QCD: Chiral Perturbation Theory}

Now let's apply these ideas to the case of QCD.  The relevant invariance
in this case is ``chiral symmetry.''   The idea of "chirality" 
is defined by the operators
\begin{equation} \Gamma_{L,R} = {1\over 2}(1\pm\gamma_5)={1\over 2}
\left( \begin{array}{c c }
1 & \mp 1 \\
\mp 1 & 1
\end{array}\right)
\end{equation}
which project left- and right-handed components of the Dirac wavefunction
via
\begin{equation} \psi_L = \Gamma_L \psi \qquad \psi_R=\Gamma_R
\psi \quad\mbox{with}\quad \psi=\psi_L+\psi_R \end{equation}
In terms of these chirality states the quark component of the QCD Lagrangian
can be written as
\begin{equation} \bar{q}(i\not\! \! D-m)q=\bar{q}_Li\not \! \! D q_L +
\bar{q}_Ri
\not\!\! D q_R -\bar{q}_L m q_R-\bar{q}_R m
q_L \end{equation}
The reason that these chirality states are called left- and right-handed can
be seen by examining helicity eigenstates of the free Dirac equation.  In the
high energy (or massless) limit we note that 
\begin{equation}
 u(p)= \sqrt{{E+m\over 2E}}
\left( \begin{array}{c }
\chi \\  {\vec{\sigma}\cdot\vec{p} \over E+m}\chi
\end{array}\right)
\stackrel{E \gg m}{\longrightarrow} \sqrt{{1\over 2}}
\left( \begin{array}{c }
\chi \\  \vec{\sigma}\cdot\hat{p} \chi
\end{array}\right)
\end{equation}

Left- and right-handed helicity eigenstates then can be identified as
\begin{equation}
u_L(p)  \sim  \sqrt{1\over 2}
\left( \begin{array}{c}
\chi \\ -\chi
\end{array} \right),\qquad
u_R(p)  \sim  \sqrt{1\over 2}
\left( \begin{array}{c}
\chi \\ \chi
\end{array} \right)
\end{equation}
But 
\begin{eqnarray}
 \Gamma_L u_L= u_L && \Gamma_R u_L=0 \nonumber \\
 \Gamma_R u_R= u_R && \Gamma_L u_R =0
\end{eqnarray}
so that in this limit chirality is identical with helicity---
\[ \Gamma_{L,R} \sim \mbox{helicity!} \]
With this background, we now return to QCD and observe that in the limit as  
$m\rightarrow0$ 
\begin{equation} {\cal L}_{\rm QCD}\stackrel{m=0}{\longrightarrow}
\bar{q}_L i \not\!\! D q_L +
\bar{q}_R i \not\!\! D q_R \end{equation}
would be invariant under {\it independent} global
left- and right-handed rotations
\begin{equation}
q_L  \rightarrow \exp (i \sum_j \lambda_j\alpha_j)
q_L,\qquad
q_R  \rightarrow \exp (i\sum_j \lambda_j \beta_j)
q_R
\end{equation}
(Of course, in this limit the heavy quark component is also invariant, but
since $m_{c,b,t} >> \Lambda_{\rm QCD}$ it would be silly to consider this as
even an approximate symmetry in the real world.)  This invariance is called
$SU(3)_L \bigotimes SU(3)_R$ or chiral $SU(3)\times SU(3)$.  Continuing
to neglect the light quark masses,
we see that in a chiral symmetric world one would expect to have 
sixteen---eight
left-handed and eight right-handed---conserved Noether currents
\begin{equation} \bar{q}_L\gamma_{\mu} {1\over 2} \lambda_i q_L \, ,
\qquad \bar{q}_R\gamma_{\mu}{1\over 2}\lambda_i
q_R \end{equation}
Equivalently, by taking the sum and difference, we would have eight 
conserved vector and
eight conserved axial vector currents
\begin{equation}
V^i_{\mu}=\bar{q}\gamma_{\mu} {1\over 2}
\lambda_i q,\qquad
A^i_{\mu}=\bar{q}\gamma_{\mu}\gamma_5
 {1\over 2} \lambda_i q
\end{equation}
In the vector case, this is just a simple generalization of 
isospin (SU(2)) invariance 
to the case of SU(3).  There exist 
{\it eight} ($3^2-1$) time-independent charges
\begin{equation} F_i=\int d^3 x V^i_0(\vec{x},t) \end{equation}
and there exist various supermultiplets of particles having 
identical spin-parity and
(approximately) the same mass in the configurations---singlet, 
octet, decuplet, 
{\it etc.} demanded by SU(3) invariance.

If chiral symmetry were realized in the conventional fashion one would
expect there also to exist corresponding nearly degenerate same spin
but {\it opposite} parity states generated
by the action of the time-independent axial charges
$F^{5}_i= \int d^3 xA^i_0(\vec{x},t)$
on these states.  Indeed since
\begin{eqnarray}
H|P\rangle &= & E_P|P\rangle \nonumber \\
H(Q_5|P\rangle)&=&Q_5(H|P\rangle)
=  E_P(Q_5|P\rangle)
\end{eqnarray}
we see that $Q_5|P\rangle$ must also be an eigenstate of the Hamiltonian
with the same eigenvalue as $|P>$, which would seem to require the existence of
parity doublets.  However, experimentally this does not appear to be the
case.  Indeed although the $J^p={1\over 2}^+$ nucleon has a 
mass of about 1 GeV, the
nearest ${1\over 2}^-$ resonance lies nearly 600 MeV higher in energy.  
Likewise in
the case of the $0^-$ pion, which has a mass of about 140 MeV, the nearest 
corresponding $0^+$ state (if it exists at all) is nearly 700 MeV or so higher 
in energy. 

The resolution of this apparent paradox is that the axial symmetry
is spontaneously broken, in which case Goldstone's theorem requires the
existence of eight massless pseudoscalar bosons, which couple derivatively
to the rest of the universe.  That way the state $Q_5|P>$ is equivalent
to $|Pa>$, where $a$ signifies one of these massless bosons, and in this
way the problem of parity doublets is avoided.  Of course, in the real
world such massless $0^-$ states do not exist.  This is because in the
real world exact chiral invariance is broken by the small quark mass terms
which we have neglected up to this point.  Thus what we have in
reality are eight
very light (but not massless) pseudo-Goldstone bosons which make up the
pseudoscalar octet.  Since such states are lighter than their other hadronic
counterparts, we have a situation wherein effective field theory can be 
applied---provided one is working at energy-momenta small compared to 
the $\sim 1$ GeV scale which is typical of hadrons, one can describe the
interactions of the pseudoscalar mesons using an effective Lagrangian.
Actually this has been known since the 1960's, where a good deal of work
was done with a {\it lowest order} effective chiral Lagrangian\cite{gg} 
\begin{equation}
 {\cal L}_2={F_\pi^2 \over 4} \mbox{Tr} (\partial_{\mu}U \partial^{\mu}
 U^{\dagger})+{m^2_{\pi}\over 4} F_\pi^2 \mbox{Tr} 
(U+U^{\dagger})\,  .\label{eq:abc}
\end{equation}
where the subscript 2 indicates that we are working at two-derivative order
or one power of chiral symmetry breaking---{\it i.e.} $m_\pi^2$.
Here $U\equiv\exp(\sum \lambda_i\phi_i/F_\pi)$, where $F_\pi=92.4$ is the pion
decay constant. This Lagrangian is {\it unique}---if we expand to 
lowest order in $\vec\phi$
\begin{eqnarray}
\mbox{Tr}\partial_{\mu} U \partial^{\mu} U^{\dagger} &=&
\mbox{Tr} {i\over F_\pi} \vec{\tau}\cdot\partial_{\mu}\vec{\phi} \times
{-i\over F_\pi}\vec{\tau}\cdot\partial^{\mu}\vec{\phi}= {2\over F_\pi^2}
\partial_{\mu}\vec{\phi}\cdot \partial^{\mu}\vec{\phi}\,  \nonumber\\
{\rm Tr}(U+U^\dagger)&=&{\rm Tr}(2-{1\over F_\pi^2}\vec{\tau}\cdot
\vec{\phi}\vec{\tau}\cdot\vec{\phi})={const.}-{2\over F_\pi^2}\vec{\phi}
\cdot\vec{\phi}
\end{eqnarray}
we reproduce the free pion Lagrangian, as required.

At the SU(3) level, including a generalized chiral symmetry breaking term,
there is even predictive power---one has
\begin{equation}
 {F_\pi^2\over 4} \mbox{Tr} \partial_{\mu} U \partial^{\mu} U^{\dagger}
=   {1\over 2} \sum_{j=1}^8 \partial_{\mu}
\phi_j\partial^{\mu}\phi_j +\cdots \nonumber\\
\end{equation}
\begin{eqnarray}
{F_\pi^2 \over 4} \mbox{Tr} 2 B_0 m ( U&+& U^{\dagger})
=  \mbox{const.}
-{1\over 2} (m_u+ m_d)B_0 \sum_{j=1}^3 \phi^2_j \nonumber\\
&-&{1\over 4} (m_u+m_d+2m_s)B_0\sum_{j=4}^7 \phi^2_j
 -{1\over 6} (m_u+m_d +4m_s)B_0\phi^2_8  +\cdots \, \nonumber\\
&&
\end{eqnarray}
where $B_0$ is a constant and $m$ is the quark mass matrix. We can
then identify the meson masses as
\begin{eqnarray}
 m^2_{\pi} & =&  2\hat{m} B_0
\nonumber \\
 m_K^2 &=& (\hat{m} +m_s) B_0 \nonumber \\
m_{\eta}^2 & =& {2\over 3} (\hat{m} + 2m_s) B_0  \, ,
\end{eqnarray}
where $\hat{m}={1\over 2}(m_u+m_d)$ is the mean light quark mass.
This system of three equations is {\it overdetermined}, and we find by simple
algebra
\begin{equation}
3m_{\eta}^2 +m_{\pi}^2 - 4m_K^2 =0 \, \, .
\end{equation}
which is the Gell-Mann-Okubo mass relation and is well-satisfied
experimentally\cite{gmo}.  Expanding to fourth order in the fields we 
also reproduce the well-known and experimentally successful
Weinberg $\pi\pi$ scattering lengths.

However, when one attempts to go beyond tree level, in order to unitarize 
the results, divergences arise and that is where the field stopped at the 
end of the 1960's.  The solution, as proposed a decade later 
by Weinberg\cite{wbp} 
and carried out by 
Gasser and Leutwyler\cite{gl}, is to absorb these 
divergences in phenomenological
constants, just as done in QED.  What is different in this case is that
the theory is nonrenormalizabile in that the forms of the divergences are
{\it different} from the terms that one started with.  That means that 
the form of the counterterms that are used to absorb these divergences 
must also be different, and Gasser and Leutwyler wrote down the most general
counterterm Lagrangian that one can have at one loop, involving {\it 
four-derivative} interactions
\begin{eqnarray}
{\cal L}_4 &  =&\sum^{10}_{i=1} L_i {\cal O}_i
= L_1\bigg[{\rm tr}(D_{\mu}UD^{\mu}U^{\dagger})
\bigg]^2+L_2{\rm tr} (D_{\mu}UD_{\nu}U^{\dagger})\cdot
{\rm tr} (D^{\mu}UD^{\nu}U^{\dagger}) \nonumber \\
 &+&L_3{\rm tr} (D_{\mu}U D^{\mu}U^{\dagger}
D_{\nu}U D^{\nu}U^{\dagger})
+L_4 {\rm tr}  (D_{\mu}U D^{\mu}U^{\dagger})
{\rm tr} (\chi{U^{\dagger}}+U{\chi}^{\dagger}
) \nonumber \\
&+&L_5{\rm tr} \left(D_{\mu}U D^{\mu}U^{\dagger}
\left(\chi U^{\dagger}+U \chi^{\dagger}\right)
\right)+L_6\bigg[ {\rm tr} \left(\chi U^{\dagger}+
U \chi^{\dagger}\right)\bigg]^2 \nonumber \\
&+&L_7\bigg[ {\rm tr} \left(\chi^{\dagger}U-
U\chi^{\dagger}\right)\bigg]^2 +L_8 {\rm tr}
\left(\chi U^{\dagger}\chi U^{\dagger}
+U \chi^{\dagger}
U\chi^{\dagger}\right)\nonumber \\
&+&iL_9 {\rm tr} \left(F^L_{\mu\nu}D^{\mu}U D^{\nu}
U^{\dagger}+F^R_{\mu\nu}D^{\mu} U^{\dagger}
D^{\nu} U \right) +L_{10} {\rm tr}\left(F^L_{\mu\nu}
U F^{R\mu\nu}U^{\dagger}\right) \nonumber\\
\end{eqnarray}
where the covariant derivative is defined via
\begin{equation}
D_\mu U=\partial_\mu U+\{A_\mu,U\}+[V_\mu,U]
\end{equation}
the constants $L_i, i=1,2,\ldots 10$ are arbitrary (not determined from chiral
symmetry alone) and
$F^L_{\mu\nu}, F^R_{\mu\nu}$ are external field strength tensors defined via
\begin{eqnarray}
F^{L,R}_{\mu\nu}=\partial_\mu F^{L,R}_\nu-\partial_\nu
F^{L,R}_\mu-i[F^{L,R}_\mu ,F^{L,R}_\nu],\qquad F^{L,R}_\mu =V_\mu\pm A_\mu .
\end{eqnarray}
Now just as in the case of QED the bare parameters $L_i$ which appear
in this Lagrangian are not physical quantities.  Instead the experimentally 
relevant (renormalized)
values of these parameters are obtained by appending to these bare values
the divergent one-loop contributions---
\begin{equation} L^r_i = L_i -{\gamma_i\over 32\pi^2}
\left[{-2\over \epsilon} -\ln (4\pi)+\gamma -1\right]\end{equation}
By comparing predictions with experiment, Gasser and Leutwyler were able 
to determine empirical numbers for each of these ten parameters.  
Typical values are shown in Table 1, together with the way in which they
were determined.
\begin{table}
\begin{center}
\begin{tabular}{l l c}\hline\hline
Coefficient & Value & Origin \\
\hline
$L_1^r$ & $0.65\pm 0.28$ & $\pi\pi$ scattering \\
$L_2^r$ & $1.89\pm 0.26$ & and\\
$L_3^r$ & $-3.06\pm 0.92$ & $K_{\ell 4}$ decay \\
$L_5^r$ & $2.3\pm 0.2$ & $F_K/F_\pi$\\
$L_9^r$ & $7.1\pm 0.3$ & $\pi$ charge radius \\
$L_{10}^r$ & $-5.6\pm 0.3$ & $\pi\rightarrow e\nu\gamma$\\
\hline\hline
\end{tabular}
\caption{Gasser-Leutwyler counterterms and the means by which
they are determined.}
\end{center}
\end{table}
 
The important question to ask at this point is why stop at order 
four derivatives?
Clearly if two-loop amplitudes from ${\cal L}_2$ or one-loop
corrections from ${\cal L}_4$ are calculated, divergences will arise which
are of six-derivative character.  Why not include these?  The answer is that
the chiral procedure represents an expansion in energy-momentum.  Corrections
to the lowest order (tree level) predictions from one 
loop corrections from ${\cal L}_2$
or tree level contributions from ${\cal L}_4$ are ${\cal
O}(E^2/\Lambda_\chi^2)$
where $\Lambda_\chi\sim 4\pi F_\pi\sim 1$ GeV is the chiral scale\cite{sca}.
Thus chiral
perturbation theory is a {\it low energy} procedure.  It is only to the extent
that the energy is small compared to the chiral scale that it makes sense to
truncate the expansion at the one-loop (four-derivative) level.  Realistically this
means that we deal with processes involving $E<500$ MeV, and, 
for such reactions the procedure is found to work very well.

In fact Gasser and Leutwyler, besides giving the form of the ${\cal O}(p^4)$ chiral
Lagrangian, have also performed the one loop integration and have written the
result in a simple algebraic form.  Users merely need to look up the result in
their paper and, despite having ten phenomenological constants, the theory is
quite predictive.  An example is shown in Table 2, where predictions are
given involving quantities which arise using just two of the 
constants---$L_9,L_{10}$.  The table also reveals at least one intruguing
problem---a solid chiral prediction, that for the charged pion 
polarizability, is possibly violated although this is not clear since there 
are three experimental results, only one of which is in disagreement.  
Clearing up this discrepancy should be a focus of future experimental work.  
Because of space limitations we shall have to be
content to stop here, but interested readers can find applications to other
systems in a number of review articles\cite{cptr}.

\begin{table}
\begin{center}
\begin{tabular}{cccc}\hline\hline
Reaction&Quantity&Theory&Experiment\\
\hline
$\pi^+\rightarrow e^+\nu_e\gamma$ & $h_V(m_\pi^{-1})$ & 0.027 
& $0.029\pm 0.017$\cite{pdg}\\
$\pi^+\rightarrow e^+\nu_ee^+e^-$ & $r_V/h_V$ & 2.6 & $2.3\pm 0.6$\cite{pdg}\\
$\gamma\pi^+\rightarrow\gamma\pi^+$ & $(\alpha_E+\beta_M)\,(10^{-4}\,{\rm fm}^3)$& 0
&$1.4\pm 3.1$\cite{anti}\\
      &$\alpha_E\,(10^{-4}\,{\rm fm}^3)$&2.8 & $6.8\pm 1.4$\cite{anti1}\\
 & & & $12\pm 20$\cite{russ}\\
 & & & $2.1\pm 1.1$\cite{slac}\\
\hline
\end{tabular}
\caption{Chiral Predictions and data in radiative pion processes.}
\end{center}
\end{table}

\section{$\chi$pt and Nucleons}

For applications invlolving nucleons it is important to note that the same ideas can
be applied within the sector of meson-nucleon interactions, although with 
a bit more difficulty.  
Again much work
has been done in this regard\cite{gss}, but there remain important 
challenges\cite{bkm}.  Writing
the lowest order chiral Lagrangian at the SU(2) level is
straightforward---
\begin{equation}
{\cal L}_{\pi N}=\bar{N}(i\not\!\!{D}-m_N+{g_A\over 2}\rlap /{u}\gamma_5)N
\end{equation}
where $g_A$ is the usual nucleon axial coupling in the chiral limit, the
covariant derivative $D_\mu=\partial_\mu+\Gamma_\mu$ is given by
\begin{equation}
\Gamma_\mu={1\over 2}[u^\dagger,\partial_\mu u]-{i\over 2}u^\dagger
(V_\mu+A_\mu)u-{i\over 2}u(V_\mu-A_\mu)u^\dagger ,
\end{equation}
and $u_\mu$ represents the axial structure
\begin{equation}
u_\mu=iu^\dagger\nabla_\mu Uu^\dagger
\end{equation}
Expanding to lowest order, we find
\begin{eqnarray}
{\cal L}_{\pi N}&=&\bar{N}(i\rlap /{\partial}-m_N)N+g_A
\bar{N}\gamma^\mu\gamma_5{1\over 2}\vec{\tau}N\cdot({i\over F_\pi}\partial_\mu\vec{\pi}
+2\vec{A}_\mu)\nonumber\\
&-&{1\over 4F_\pi^2}\bar{N}\gamma^\mu\vec{\tau}N\cdot\vec{\pi}\times
\partial_\mu\vec{\pi}+\ldots
\end{eqnarray}
which yields the Goldberger-Treiman relation,
connecting strong and weak couplings of the nucleon system\cite{gt}
\begin{equation}
F_\pi g_{\pi NN}=m_N g_A
\end{equation}
Using the present best values for these quantities, we find

\begin{equation}
92.4 \mbox{MeV}\times 13.05 =1206 \mbox{MeV}\quad\mbox{vs.}\quad 1189 \mbox{MeV}
= 939\mbox{MeV}\times 1.266
\end{equation}
and the agreement to better than two percent strongly confirms the validity
of chiral symmetry in the nucleon sector.  Actually the Goldberger--Treiman relation
is only strictly true at the unphysical point $g_{\pi NN}(q^2=0)$ 
and one {\it expects}
about a ~1\% discrepancy to exist.  An interesting "wrinkle" in this regard
is the use of the so-called Dashen-Weinstein relation, which accounts
for lowest order SU(3)
symmetry breaking effects, to predict this discrepancy in terms of corresponding numbers in the
strangeness changing sector\cite{dw}.

  However, any realistic approach must also involve loop calculations 
as well as the use of a Foldy-Wouthuysen transformation in order to assure
proper power counting.  This approach goes under the name of heavy baryon
chiral perturbation theory (HB$\chi$pt) and interested readers can find a 
compendium of such results in the review article\cite{bkm}.  
For our purposes we shall have to be content to examine just one 
application---measurement of 
the {\it generalized} proton {\it polarizability} 
via virtual Compton scattering.  First recall from section 3 
the concept of polarizability as the constant of
proportionality between an applied electric or magnetizing field and the
resultant induced electric or magnetic dipole moment---
\begin{equation}
\vec{p}=4\pi\alpha_E\vec{E},\qquad \vec{\mu}=4\pi\beta_M\vec{H}
\end{equation}
The corresponding interaction energy is
\begin{equation}
E=-{1\over 2}4\pi\alpha_EE^2-{1\over 2}4\pi\beta_MH^2
\end{equation}
which, upon quantization, leads to a proton Compton scattering cross section
\begin{eqnarray}
{d\sigma\over d\Omega}&=&\left({\alpha_{em}\over m}\right)^2\left({\omega'\over
\omega}\right)^2[{1\over 2}
(1+\cos^2\theta)\nonumber\\
&-&{m\omega\omega'\over \alpha_{em}}[{1\over
2}(\alpha_E+\beta_M)(1+\cos\theta)^2
+{1\over 2}(\alpha_E-\beta_M)(1-\cos\theta)^2+\ldots].\label{eq:sss}
\end{eqnarray}
It is clear from Eq.(\ref{eq:sss})
that, from careful measurement of the differential scattering cross section,
extraction of these structure dependent polarizability terms is possible
provided that
\begin{itemize}
\item [i)] the energy is large enough that these terms are significant compared to the
leading
Thomson piece and 
\item [ii)] that the energy is not so large that higher order
corrections become important 
\end{itemize}
and this has been accomplished recently at
SAL and MAMI, yielding\cite{protpol}
\begin{equation}
\alpha_E^{exp}=(12.1\pm 0.8\pm 0.5)\times 10^{-4}\mbox{fm}^3,\qquad
\beta_M^{exp}=(2.1\mp 0.8\mp 0.5)\times
10^{-4}\mbox{fm}^3
\end{equation}
A chiral one loop calculation has also been performed by Bernard, Kaiser, and
Meissner and yields a result in good agreement with these 
measurements\cite{bkmt}
\begin{equation}
\alpha_E^{theo}=10\beta_M^{theo}={5e^2g_A^2\over 384\pi^2F_\pi^2m_\pi}=
12.2\times 10^{-4}\mbox{fm}^3
\end{equation}
The idea of {\it generalized} polarizability can be understood from the
analogous venue of electron scattering wherein measurement of the charge
form factor as a function of $\vec{q}^2$ leads, when Fourier transformed,
to a picture of the {\it local} charge density within the system.  In the
same way the virtual Compton scattering process---$\gamma^*+p\rightarrow
\gamma+p$ can provide a measurement of the $\vec{q}^2$-dependent 
electric and magnetic polarizabilities, whose Fourier transform
provides a picture of the {\it local
polarization density} within the proton.  On the theoretical side our group 
has performed a one loop HB$\chi$pt calculation and has produced a closed from
expression for the predicted polarizabilities\cite{hemm}
 \begin{eqnarray}
\bar{\alpha}^{(3)}_E(\bar{q})&=& \frac{e^2 g_{A}^2 m_\pi}{64\pi^2
F_{\pi}^2}\;\frac{4+2\frac{\bar{q}^2}{m_{\pi}^2}-\left(8-2\frac{\bar{q}^2}{m_{\pi}^2}
-\frac{\bar{q}^4}{m_{\pi}^4}\right)\frac{m_\pi}{\bar{q}}\arctan\frac{\bar{q}}{2
m_{\pi}}}{\bar{q}^2\left(4+\frac{\bar{q}^2}{m_{\pi}^2}\right)}
\; , \nonumber\\
\bar{\beta}^{(3)}_M(\bar{q})&=& \frac{e^2 g_{A}^2 m_\pi}{128\pi^2
F_{\pi}^2}\;\frac{-\left(4+2\frac{\bar{q}^2}{m_{\pi}^2}\right)+\left(8+6\frac{
\bar{q}^2}{m_{\pi}^2}+\frac{\bar{q}^4}{m_{\pi}^4}\right)\frac{m_\pi}{\bar{q}}\arctan
\frac{\bar{q}}{2 m_{\pi}}}{\bar{q}^2\left(4+\frac{\bar{q}^2}{m_{\pi}^2}\right)}
\; . \label{eq:bq}
\end{eqnarray}
In the electric case the structure is about what would be expected---a gradual
falloff of $\alpha_E(\bar{q})$ from the real photon point with scale
$r_p\sim 1/m_\pi$.  However, the magnetic generalized polarizability is
predicted to {\it rise} before this general falloff occurs---chiral
symmetry requires the presence of both a paramagnetic and a diamagnetic
component to the proton.  Both predictions have received some
support in a soon to be announced (and tour de force) MAMI measurement
at $\bar{q}=600$ MeV\cite{mami}.  However, 
since parallel kinematics were employed in the experiment
the desired generalized polarizabilities had to be identified on top
of an enormous Bethe-Heitler background.  A Bates measurement, 
to be performed by the OOPS collaboration next spring, will take 
place at $\bar{q}=240$ MeV
and will use the cababilities of the OOPS detector system to provide
a 90 degree out of plane measurement, which will be {\it much} less
sensitive to the Bethe-Heitler blowtorch.  We anxiously await the results.

\section{Summary}
Above we have discussed some of the consequences of the feature that
the $SU(3)_L\times SU(3)_R$ chiral symmetry of QCD is broken spontaneously 
in the axial sector,
implying the existence of eight pseudo-Goldstone bosons which, because they 
are considerably lighter than the remaining hadronic spectrum, can be
treated via an effective field theory---chiral perturbation theory.  The 
predictions arising from such calculations are rigorous consequences of
the underlying chiral symmetry of QCD and are subject to experimental
tests using TAPS or other detectors at low energy machines.  A taste of
the predictive power is given above, but readers seeking a more 
substantial meal can find extensive 
summaries in various other sources\cite{cptr}.

\begin{center}
{\bf Acknowledgement}
\end{center}

It is a pleasure to thank the organizers for inviting me to participate in
this workshop and for hospitality in Rez.  
This work was supported in part by the National Science Foundation.


\begin{thebibliography}{99}
\bibitem{ssb} J. Svardiere, Am. J. Phys. {\bf 51} (1983) 1016.
\bibitem{afp} See, {\it e.g.}, B.R. Holstein, Am. J. Phys. {\bf 61}
(1993) 142 and references therein; R. Jackiw, in {\bf M.A.B. Memorial Volume},
ed. A. Ali and P. Hoodboy, World Scientific. Singapore (1991).
\bibitem{eftr} See, {\it e.g.} A. Manohar, "Effective Field Theories," in
{\bf Schladming 1966: Perturbative and Nonperturbative 
Aspects of Quantum Field
Theory}, hep-ph/9606222; D. Kaplan, "Effective Field 
Theories," in Proc. 7th Summer
School in Nuclear Physics, nucl-th/9506035,; H. Georgi, "Effective 
Field Theory," in
Ann. Rev. Nucl Sci. {\bf 43} (1995) 209.
\bibitem{skb} B.R. Holstein, Am. J. Phys. {\bf 67} (1999) 422.
\bibitem{feyn} A corresponding classical physics discussion is given 
in R.P Feynman,
R.B. Leighton, and M. Sands, {\bf The Feynman Lecures on Physics}, 
Addison-Wesley, Reading, MA, (1963) Vol. I, Ch. 32. 
\bibitem{gg} S. Gasiorowicz and D.A. Geffen, Rev. Mod. Phys. {\bf 41} 
(1969) 531.
\bibitem{gmo} M. Gell-Mann, CalTech Rept. {\bf CTSL-20} (1961); S. Okubo,
Prog. Theo. Phys. {\bf 27} (1962) 949.
\bibitem{wbp} S. Weinberg, Physica {\bf A96} (1979) 327.
\bibitem{gl} J. Gasser and H. Leutwyler, Ann. Phys. (NY) {\bf 158} 
(1984) 142; Nucl. Phys. {\bf B250} (1985) 465.  
\bibitem{sca} A. Manohar and H. Georgi, Nucl. Phys. {\bf B234} (1984) 189;
J.F. Donoghue, E. Golowich and B.R. Holstein, Phys. Rev. {\bf D30}
(1984) 587.
\bibitem{pdg} Particle Data Group, Phys. Rev. {\bf D54} (1996) 1.
\bibitem{anti} Yu. M. Antipov et al., Z. Phys. {\bf C26} (1985) 495. 
\bibitem{anti1} Yu. M. Antipov et al., Phys. Lett. {\bf B121} (1983) 445.
\bibitem{russ} T.A. Aibergenov et al., Czech. J. Phys. {\bf 36} (1986) 948.
\bibitem{slac} D. Babusci et al., Phys. Lett. {\bf B277} (1992) 158.
\bibitem{cptr} See, {\it e.g.} B.R. Holstein, Int.J. Mod. Phys. {\bf A7} 
(1993) 7873; H. Leutwyler, in {\bf Perspectives in the Standard Model}, eds. 
R.K. Ellis, C.T. Hill, and J.D. Lykken, World Scientific, Singapore (1992);
J. Gasser, in Advanced School on Effective Theories, eds. F. Cornet and M.J.
Herrero, World Scientific, Singapore (1997); H. Leutwyler, in {\bf Selected 
Topics in Nonperturbative QCD}, eds. A. DiGiacomo and D. Diakonov, IOS
Press, Amsterdam (1996).
\bibitem{gss} J. Gasser, M. Sainio, and A. Svarc, Nucl. Phys. {\bf B307}
(1988) 779. 
\bibitem{bkm} V. Bernard, N. Kaiser, and U.-G. Meissner, Int. J. Mod. Phys. 
{\bf E4} (1995) 193.
\bibitem{gt} M. Goldberger and S.B. Treiman, Phys. Rev. {\bf 110} (1958) 1478.
\bibitem{dw} R. Dashen and M. Weinstein, Phys. Rev. {\bf 188} (1969) 2330; 
B.R. Holstein, "Nucleon Axial Matrix Elements," 
Few-Body Systems Suppl. {\bf 11} (1999) 116; J.L. Goity, R. Lewis, 
and M. Schvelinger, "The Goldberger-Treiman Discrepancy in SU(3)," Phys.
Lett. {\bf B454} (1999) 115.
\bibitem{protpol} F.J. Federspiel et al., Phys. Rev. Lett. {\bf 67} (1991)
1511; A. L. Hallin et al., Phys. Rev. {\bf C48} (1993) 1497;
A. Zieger et al., Phys. Lett. {\bf B278} (1992) 34; B.R. MacGibbon et
al., Phys. Rev. {\bf C52} (1995) 2097.
\bibitem{bkmt} V. Bernard, N. Kaiser, and U.-G. Meissner, Phys. Rev.
Lett. {\bf 67} (1991) 1515.
\bibitem{hemm} T.R. Hemmert, B.R. Holstein, G. Knoechlein, and D.
Drechsel, hep-ph/9910036.
\bibitem{mami} S. Kerhoas et al., Few Body Syst. Supp. {\bf 10} (1999) 523.

\end{thebibliography}
\end{document}